\documentclass[aps,prl,showpacs,noshowkeys,amsmath,amssymb,amsfonts,superscriptaddress,reprint]{revtex4-1}

\usepackage{graphicx}
\usepackage{bm}
\usepackage{physics}
\usepackage{mathtools}

\usepackage[colorlinks=true,linkcolor=blue,citecolor=blue,urlcolor=blue]{hyperref}
\usepackage[all]{hypcap}
\usepackage[nameinlink,capitalise]{cleveref}
\begin{document}
\title{Active Fingering Instability in Tissue Spreading}
\author{Ricard Alert}
\email{ricard.alert@princeton.edu}
\altaffiliation{Present address: Princeton Center for Theoretical Science and Lewis-Sigler Institute for Integrative Genomics, Princeton University, Princeton NJ 08544, USA}
\affiliation{Departament de F\'{i}sica de la Mat\`{e}ria Condensada, Universitat de Barcelona, Av. Diagonal 647, 08028 Barcelona, Spain}
\affiliation{Universitat de Barcelona Institute of Complex Systems (UBICS), Universitat de Barcelona, Barcelona, Spain}
\author{Carles Blanch-Mercader}
\affiliation{Laboratoire Physico Chimie Curie, Institut Curie, PSL Research University, CNRS, 26 rue d'Ulm, 75005 Paris, France}
\affiliation{Department of Biochemistry, Faculty of Sciences, University of Geneva, 30, Quai Ernest-Ansermet, 1205 Gen\`{e}ve, Switzerland}
\author{Jaume Casademunt}
\affiliation{Departament de F\'{i}sica de la Mat\`{e}ria Condensada, Universitat de Barcelona, Av. Diagonal 647, 08028 Barcelona, Spain}
\affiliation{Universitat de Barcelona Institute of Complex Systems (UBICS), Universitat de Barcelona, Barcelona, Spain}
\date{\today}

\begin{abstract}
During the spreading of epithelial tissues, the advancing tissue front often develops finger-like protrusions. Their resemblance to traditional viscous fingering patterns in driven fluids suggests that epithelial fingers could arise from an interfacial instability. However, the existence and physical mechanism of such a putative instability remain unclear. Here, based on an active polar fluid model for epithelial spreading, we analytically predict a generic instability of the tissue front. On the one hand, active cellular traction forces impose a velocity gradient that leads to an accelerated front, which is, thus, unstable to long-wave\-length perturbations. On the other hand, contractile intercellular stresses typically dominate over surface tension in stabilizing short-wave\-length perturbations. Finally, the finite range of hydrodynamic interactions in the tissue selects a wavelength for the fingering pattern, which is, thus, given by the smallest between the tissue size and the hydrodynamic screening length. Overall, we show that spreading epithelia experience an active fingering instability based on a simple kinematic mechanism. Moreover, our results underscore the crucial role of long-range hydrodynamic interactions in the dynamics of tissue morphology.
\end{abstract}

\maketitle

\begin{figure}[bt]
\includegraphics[width=\columnwidth]{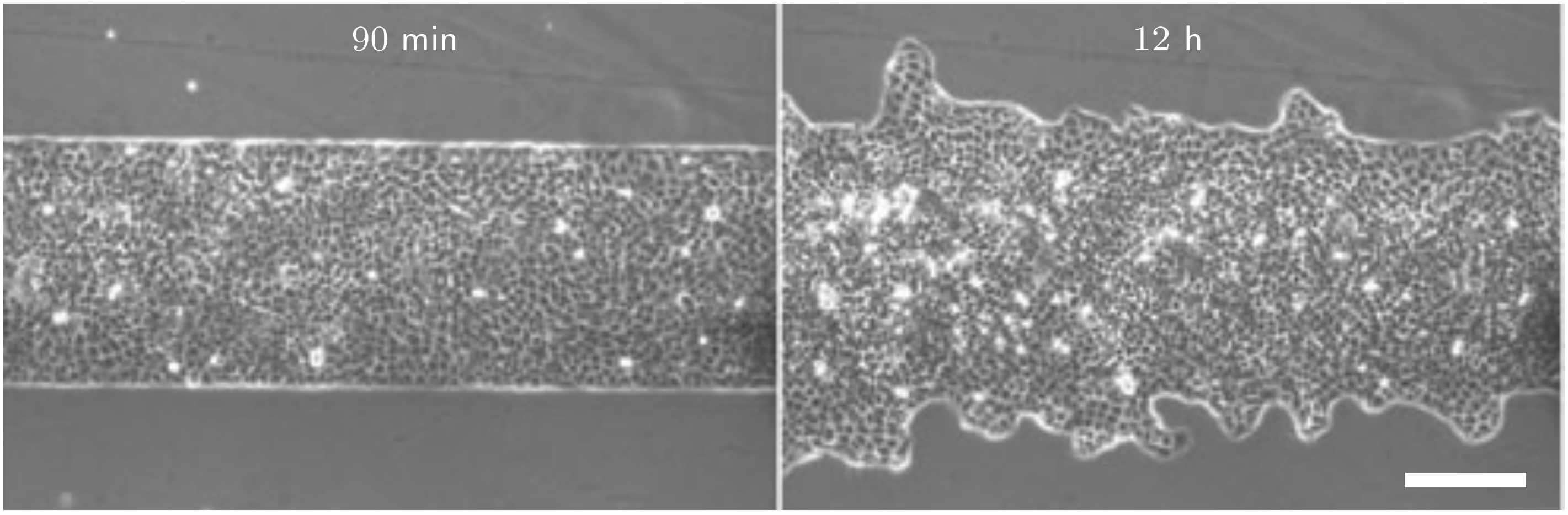}
\caption{Fingering in epithelial spreading. Scale bar, $200$ $\mu$m. Adapted from \cite{Poujade2007} with permission from Pascal Silberzan.} \label{fig fingering-epithelium}
\end{figure}

The spreading of epithelial monolayers by collective cell migration is crucial for tissue morphogenesis, wound healing, and tumor progression. Both in vivo and in vitro, multicellular protrusions called epithelial fingers often appear at the front of spreading tissues (\cref{fig fingering-epithelium}) \cite{Vedula2013a,Saw2014,Hakim2017a,Ladoux2017a,Omelchenko2003,Poujade2007,Petitjean2010,Reffay2011,Klarlund2012,Reffay2014,Vishwakarma2018}. This \emph{epithelial fingering} resembles the viscous fingering that occurs via the Saffman-Taylor instability when a viscous fluid displaces a more viscous one \cite{Saffman1958,Casademunt2004}. However, the mechanisms of these two phenomena must be different because epithelial monolayers are more viscous than the fluid that they displace. Hence, several models of epithelial fingering have been proposed \cite{Hakim2017a}.

To induce finger formation, some models directly implement \emph{leader cells} with a distinct behavior, either via special particles \cite{Sepulveda2013} or via a dependence of the magnitude of cell motility forces on the curvature of the tissue front \cite{Mark2010,Tarle2015}. Other models recapitulate epithelial fingering by introducing alignment between cell motility forces and the tissue velocity field \cite{Basan2013}. These models predict a moving front to be stable and a non-moving front to exhibit an instability with an unbounded growth rate for a number of finite wavelengths \cite{Zimmermann2014a,Nesbitt2017a}. Fingers were also observed in the numerical solution of other continuum models of spreading epithelia, either treated as active polar fluids \cite{Lee2011a} or as active nematics with cell proliferation \cite{Doostmohammadi2015}. Recently, fingering was also found in a parameter range of an active vertex model \cite{Barton2017}. Finally, interface undulations can emerge from the coupling of chemotactic fields to the mechanics of epithelial spreading \cite{Ouaknin2009,Salm2012,Kopf2013,BenAmar2016}.

Despite the many efforts, the physical mechanism of the fingering instability in epithelia remains a matter of debate. Here, we address this problem by means of a continuum active polar fluid model for epithelial spreading. The model includes hydrodynamic interactions through the tissue, and it implements neither leader-cell behavior nor alignment between cellular traction forces and the flow field. Yet, we analytically predict a long-wave\-length instability of the moving front that explains epithelial fingering. The instability is based on a generic kinematic mechanism, namely the front acceleration associated to a fixed velocity gradient. In spreading epithelia, the velocity gradient is imposed by active traction forces at the edge of the viscous cell monolayer. The fastest-growing mode has a finite wavelength, typically a few hundreds of micrometers, consistent with the measured finger spacing \cite{Vishwakarma2018}. This characteristic wavelength is selected by the long-range hydrodynamic interactions in the tissue, which are either limited by the tissue size or screened by cell-substrate friction forces. The model also shows that intercellular contractility stabilizes short-wave\-length perturbations of the tissue boundary. The stabilizing effect of contractility is typically stronger than that of tissue surface tension. Globally, our analysis shows how, as a result of the flows induced by the traction force field, a morphological instability may naturally take place in a spreading cell monolayer. Leader cells could then appear upon the onset of the instability, influencing finger development.

\paragraph{Model.---} We base our analysis on a continuum active polar fluid model of epithelial spreading, which is thus described in terms of a polarity field $\vec{p}\left(\vec{r},t\right)$ and a velocity field $\vec{v}\left(\vec{r},t\right)$ \cite{Blanch-Mercader2017,Perez-Gonzalez2019,Alert2018c}. We neglect cell proliferation and the bulk elasticity of the monolayer, which eventually limit the spreading process \cite{Poujade2007,Serra-Picamal2012,Basan2013,Recho2016,Yabunaka2017a}. Tissue spreading is primarily driven by the traction forces exerted by cells close to the monolayer edge, which polarize perpendicularly to the edge by extending lamellipodia towards free space. In contrast, the inner region of the monolayer remains essentially unpolarized, featuring much weaker and transient traction forces \cite{Blanch-Mercader2017,Perez-Gonzalez2019}. Hence, we take a free energy for the polarity field that favors the unpolarized state $p=0$ in the bulk, with a restoring coefficient $a>0$, and we impose a normal and maximal polarity as a boundary condition at the tissue edge. In addition, the polar free energy includes a cost for polarity gradients, with $K$ the Frank constant of nematic elasticity in the one-constant approximation \cite{deGennes-Prost}. Altogether,
\begin{equation} \label{eq polar-free-energy}
F=\int \left[\frac{a}{2}p_\alpha p_\alpha + \frac{K}{2}\left(\partial_\alpha p_\beta\right)\left(\partial_\alpha p_\beta\right)\right] \dd^3\vec{r}.
\end{equation}
We assume that the polarity field is set by flow-independent mechanisms, so that it follows a purely relaxational dynamics, and that it equilibrates fast compared to the spreading dynamics \cite{SM}. Hence, $\delta F/\delta p_\alpha=0$, which yields
\begin{equation} \label{eq polarity-field}
L_c^2 \nabla^2 p_\alpha=p_\alpha,
\end{equation}
where $L_c=\sqrt{K/a}$ is the characteristic length with which the polarity modulus decays from $p=1$ at the monolayer edge to $p=0$ at the center.

Then, force balance imposes
\begin{equation} \label{eq force-balance}
\partial_\beta \sigma_{\alpha\beta} + f_\alpha=0,
\end{equation}
where $\sigma_{\alpha\beta}$ is the stress tensor of the monolayer, and $f_\alpha$ is the external force density acting on it. Since tissue spreading occurs over time scales of several hours (\cref{fig fingering-epithelium}), we neglect the elastic response of the tissue \cite{SM}. Thus, we relate tissue forces to the polarity and velocity fields via the following constitutive equations for an active polar fluid \cite{Oriola2017} (see discussion and justification in Ref. \cite{SM}):
\begin{subequations}
\begin{align}
\sigma_{\alpha\beta} &= \eta\left(\partial_\alpha v_\beta + \partial_\beta v_\alpha\right) - \zeta p_\alpha p_\beta,\\
f_\alpha &= -\xi v_\alpha + \zeta_i p_\alpha.
\end{align}
\end{subequations}
Here, $\eta$ is the effective monolayer viscosity, and $\xi$ is the cell-substrate friction coefficient. Respectively, $\zeta<0$ is the active stress coefficient accounting for the contractility of polarized cells, and $\zeta_i>0$ is the contact active force coefficient accounting for the maximal traction stress exerted by polarized cells on the substrate, $T_0=\zeta_i h$, with $h$ the monolayer height.

\begin{figure}[tb]
\begin{center}
\includegraphics[width=\columnwidth]{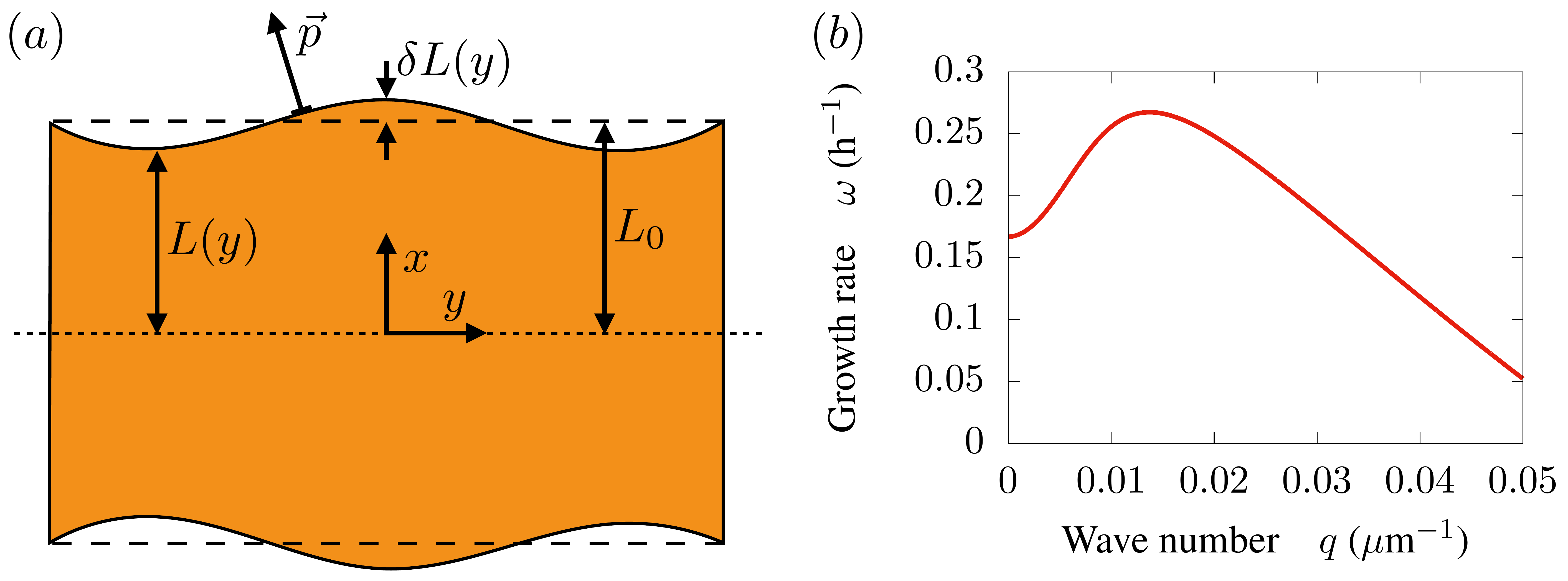}
\caption{Instability of the monolayer front. (a) Sketch of the peristaltic perturbations. Dashed lines indicate the flat, unperturbed interface. The dotted line indicates the symmetry axis of the monolayer. (b) Growth rate of the perturbations. Parameter values are in \cref{t parameters}.} \label{fig instability}
\end{center}
\end{figure}

\paragraph{Stability of the tissue front.---} To study the stability of the advancing front, we consider a rectangular monolayer typical of in vitro experiments (\cref{fig fingering-epithelium}). Thus, the reference state is the flat front solution with $\vec{p}=p_x^0(x)\, \hat{x}$ and $\vec{v}=v_x^0(x)\,\hat{x}$ \cite{SM} (dashed lines in \cref{fig instability}a). In addition to a maximal normal polarity at the edges, we impose stress-free boundary conditions. For an interface of arbitrary shape, $\vec{p}\left(x=\pm L\right)=\hat{n}_{\pm}$, and $\left.\bm{\sigma}\cdot\hat{n}_{\pm}\right|_{x=\pm L}=\vec{0}$, respectively, where $\hat{n}_\pm$ is the normal unit vector of the top and bottom interfaces. The tissue width $L$ changes according to $\dd L/\dd t=\left.\vec{v}\cdot\hat{n}\right|_{x=L}$. Then, motivated by experimental observations (\cref{fig fingering-epithelium}), we introduce peristaltic small-amplitude perturbations of the flat interface, namely those that modify the monolayer width (\cref{fig instability}a): $L(y)=L_0+\delta L(y)$. From a linear stability analysis \cite{SM}, we obtain that the growth rate $\omega(q)$ of such perturbations is always real, so that no oscillatory behavior is expected. However, the growth rate evidences a long-wave\-length instability of the monolayer front. Moreover, the fastest-growing perturbation has a finite wavelength (\cref{fig instability}b). In the following, we analyze the contribution of the different forces to the instability, which allows us to single out its physical mechanism.

\begin{figure}[tb]
\begin{center}
\includegraphics[width=\columnwidth]{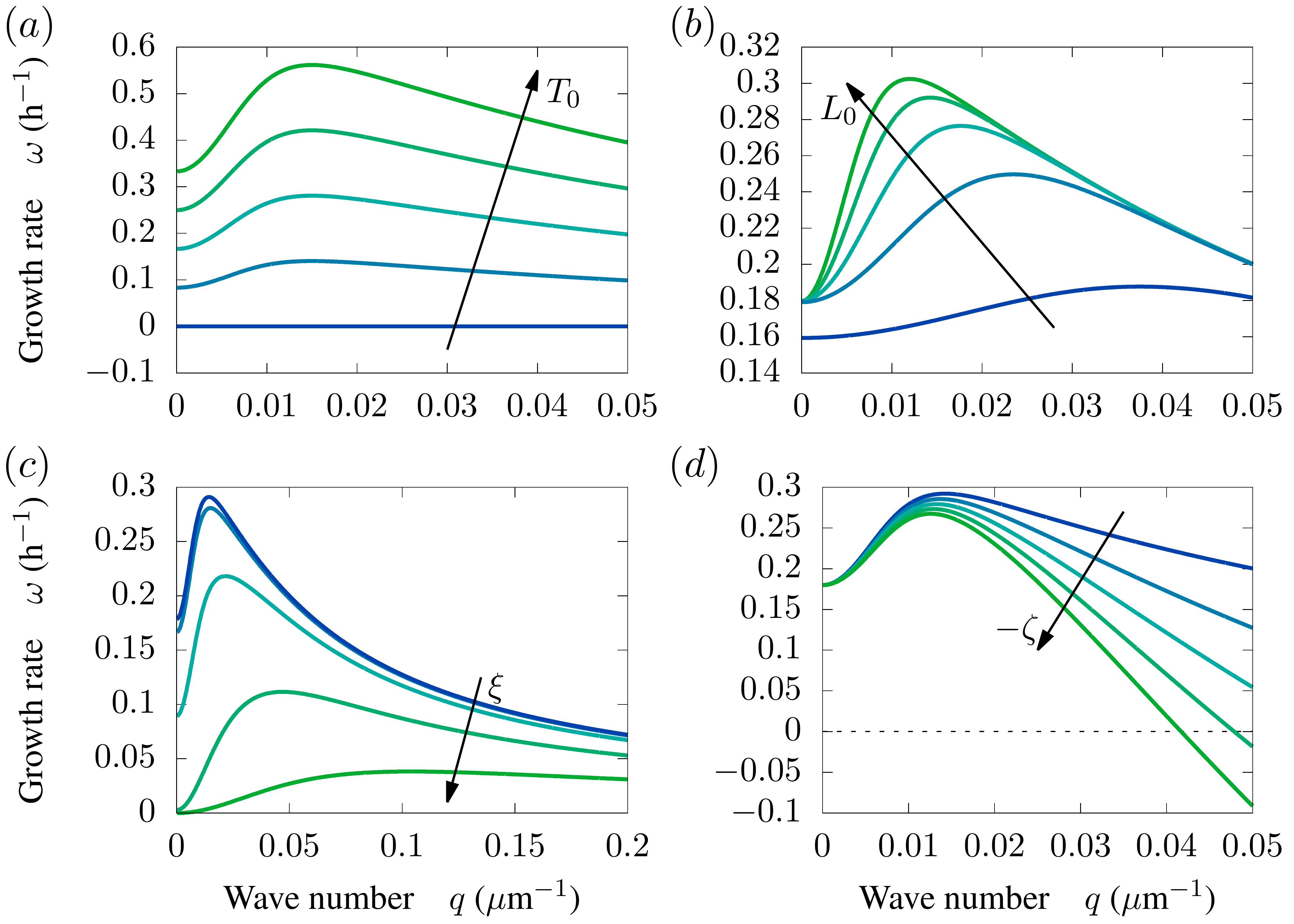}
\caption{Contributions to the instability. Growth rates of shape perturbations varying the values of different model parameters. Excluding the varied parameter, other parameter values are in \cref{t parameters} except for $\xi,\zeta\rightarrow 0$. (a) Traction forces completely destabilize the monolayer front. For this plot, $T_0=0,0.25,0.5,0.75,1$ kPa. (b) Long-range transmission of viscous stresses selects the fastest-growing mode. For this plot $L_0=50,100,150,200,250$ $\mu$m. (c) Cell-substrate friction screens hydrodynamic interactions to limit the wavelength of the fingering pattern. For this plot, $\xi=10,10^2,10^3,10^4,10^5$ Pa$\cdot$s/$\mu$m$^2$. (d) Contractility stabilizes short-wave\-length perturbations of the monolayer front. For this plot, $-\zeta=0,10,20,30,40$ kPa.} \label{fig origins}
\end{center}
\end{figure}

\paragraph{Traction forces.---} We first consider a limit case with neither intercellular contractility nor cell-substrate friction, $\zeta,\xi\rightarrow 0$ \cite{Blanch-Mercader2017}. In addition, we also consider that the width of the polarized boundary layer of cells is much smaller than the total tissue width, $L_c\ll L_0$, which is generally the case in experiments \cite{Blanch-Mercader2017,Perez-Gonzalez2019}. In this limit, since active forces are concentrated at the narrow boundary layer, most of the tissue behaves as a passive viscous fluid, for which $\partial_x\sigma_{xx}\approx 0$ and $\sigma_{xx}\approx 2\eta\, \dd v_x/\dd x$. Therefore, the stress is uniform throughout most of the tissue, with a value given by the stress accumulated across the boundary layer, namely $\sigma_{xx}\approx T_0 L_c/h$. Consequently, the velocity gradient is also fixed and uniform, and hence the spreading velocity $v_x(L_0)=V_0=\dd L_0/\dd t$ reads
\begin{equation} \label{eq front-acceleration}
V_0\approx \frac{T_0 L_c}{2\eta h}L_0 \equiv\frac{L_0}{\tau},
\end{equation}
where we have used that $v_x(0)=0$. This result means that, due to the sole action of a constant traction force, the flat front accelerates, consistent with measurements \cite{Rosen1980,Poujade2007} and with the size dependence of tissue wetting \cite{Perez-Gonzalez2019}. Consequently, the $q=0$ perturbation mode is unstable, $\omega\left(q=0\right)=\tau^{-1}>0$, since any uniform displacement of the advancing front makes it depart from its original velocity. Thus, the instability mechanism is kinematic in nature: The advanced regions of the front move faster than the trailing regions (\cref{fig instability}a). Thereby, traction forces contribute to destabilize perturbations of all wavelengths (\cref{fig origins}a). Therefore, since other forces such as surface tension stabilize short-wave\-length perturbations, the tissue front experiences a long-wave\-length instability.

\begin{table}[bt]
\begin{center}
\begin{tabular}{clc}
Symbol&Description&Estimate\\\hline
$L_0$&monolayer half-width&$200$ $\mu$m\\
$h$&monolayer height&$5$ $\mu$m \cite{Trepat2009,Perez-Gonzalez2019}\\
$L_c$&nematic length&$25$ $\mu$m \cite{Blanch-Mercader2017,Perez-Gonzalez2019}\\
$T_0$&maximal traction&$0.5$ kPa \cite{Blanch-Mercader2017,Perez-Gonzalez2019}\\
$-\zeta$&intercellular contractility&$20$ kPa \cite{Perez-Gonzalez2019}\\
$\xi$&friction coefficient&$100$ Pa$\cdot$s/$\mu$m$^2$ \cite{Cochet-Escartin2014}\\
$\eta$&monolayer viscosity&$25$ MPa$\cdot$s \cite{Blanch-Mercader2017,Perez-Gonzalez2019}\\
$\lambda$&hydrodynamic screening length&$0.5$ mm ($\sqrt{\eta/\xi}$)
\end{tabular}
\end{center}
\caption{Estimates of model parameters.} \label{t parameters}
\end{table}

\begin{figure}[tb]
\begin{center}
\includegraphics[width=\columnwidth]{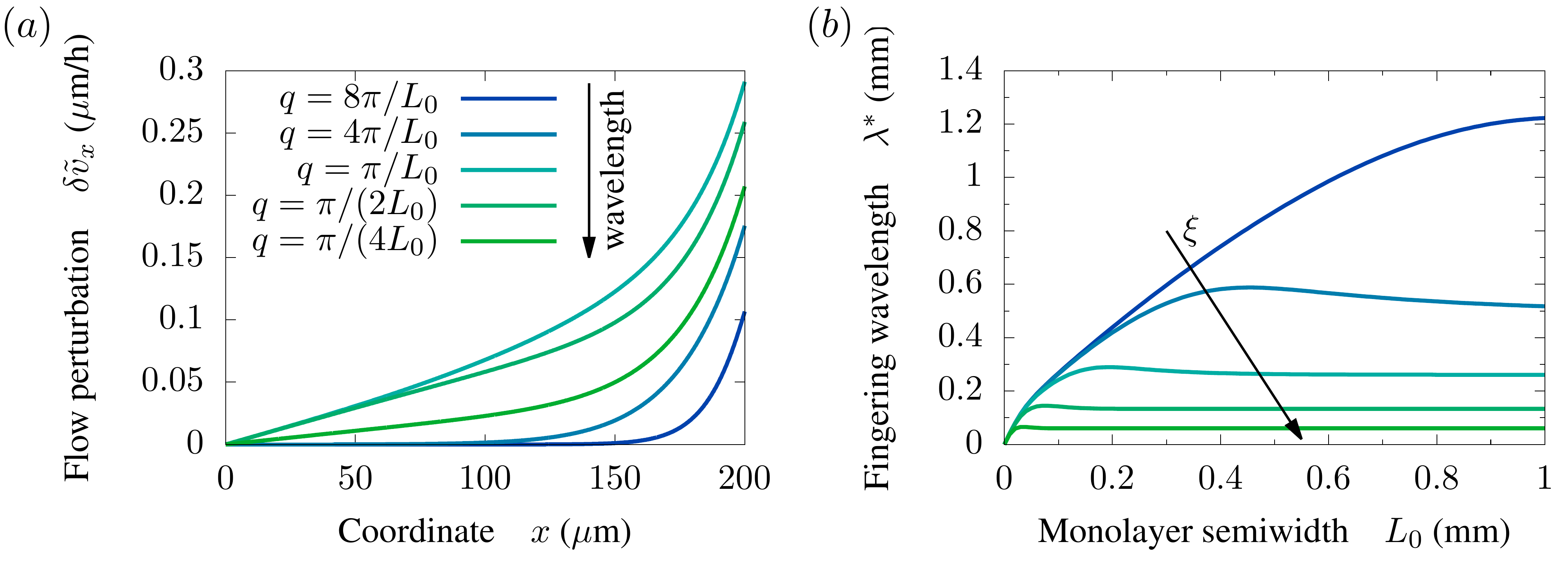}
\caption{Screening of tissue flows. (a) Flow perturbations induced by short-wave\-length shape perturbations ($q>\pi/L_0$) penetrate a distance given by their wavelength. Thus, the interfacial velocity perturbation increases with wavelength. In contrast, the penetration of flow perturbations induced by long-wave\-length shape perturbations ($q>\pi/L_0$) is limited by the tissue width $2L_0$, which entails a decrease of the interfacial velocity perturbation. Consequently, shape pertubations with a wavelength that matches the monolayer width ($q\sim \pi/L_0$) feature the fastest growth (\cref{fig origins}b). Parameter values are in \cref{t parameters} except for $\xi,\zeta\rightarrow 0$. (b) The wavelength of the fastest-growing mode, $\lambda^*$, is proportional to the monolayer semiwidth $L_0$ if $L_0\lesssim \lambda$, with $\lambda=\sqrt{\eta/\xi}$ the hydrodynamic screening length. For wider monolayers, the selected wavelength is size-independent, becoming proportional to $\lambda$. Parameter values are in \cref{t parameters} except for $\zeta\rightarrow 0$, and $\xi=10,10^2,10^3,10^4,10^5$ Pa$\cdot$s/$\mu$m$^2$.} \label{fig screening}
\end{center}
\end{figure}

\paragraph{Viscous stresses.---} The kinematic mechanism explains why long-wave\-length modes are unstable. However, it does not explain why the most unstable mode occurs at a finite wavelength (\cref{fig origins}a). In fact, the existence of a peak in the growth rate is due to the transmission of viscous stress across the monolayer. In the so-called wet limit $\lambda=\sqrt{\eta/\xi}\gg L_0$, corresponding to $\xi\rightarrow 0$, viscous stresses transmit through the entire monolayer. Thus, a given perturbation of the front generates a flow perturbation that penetrates a distance of the order of its wavelength, $\pi/q$, into the monolayer. At the monolayer edge, the stress-free boundary condition imposes $\delta \sigma_{xx}(\pm L) = \mp \partial_x \sigma_{xx}^0 (\pm L_0)\;\delta L$. Hence, since $\delta\sigma_{xx}=2\eta\,\partial_x \delta v_x$ in the absence of contractility ($\zeta\rightarrow 0$), the gradient of the velocity perturbation profile is fixed at the boundary, being positive (negative) for advanced (trailing) regions of the tissue front. Therefore, flow perturbations further destabilize the flat front in a wavelength-dependent manner.

For short wavelengths, $\pi/q<L_0$, the penetration distance of flow perturbations is shorter than the tissue width. Thus, since the slope of velocity perturbations at the interface is fixed, the longer the wavelength, the larger the interfacial velocity perturbation (\cref{fig screening}a). Hence, the growth rate increases with the wavelength (\cref{fig origins}a). In contrast, for long wavelengths, $\pi/q>L_0$, the penetration distance of flow perturbations is longer than the tissue width. Thus, in this case, the decay of flow perturbations becomes nearly linear, with a slope that decreases with increasing wavelength. Consequently, perturbations of longer wavelength feature a smaller interfacial velocity perturbation (\cref{fig screening}a), and hence they are less unstable (\cref{fig origins}a). In conclusion, in the absence of cell-substrate friction forces, the finite width of the monolayer limits the range of hydrodynamic interactions in the tissue, thus giving rise to the peak in the growth rate at $q^*\sim \pi/L_0$ (\cref{fig origins}b).

\paragraph{Cell-substrate friction forces.---} Cell-substrate friction screens the transmission of viscous stresses over distances larger than $\lambda=\sqrt{\eta/\xi}$. Consequently, the peak of the growth rate occurs at $q^*\sim \pi/\lambda$ if $\lambda\lesssim L_0$ (\cref{fig origins}c). Thus, for sufficiently strong friction, the fingering wavelength is given by the hydrodynamic screening length $\lambda$ instead of the monolayer width $L_0$ (\cref{fig screening}b). We estimate $\lambda \sim 0.5$ mm (\cref{t parameters}). Therefore, the crossover from a viscosity-dominated to a friction-dominated regime of monolayer spreading and fingering should be observable in usual in vitro experiments.

\paragraph{Surface tension.---} The monolayer edge presents a surface tension $\gamma$, namely the work per unit area required to expand it. For a curved interface, surface tension gives rise to a normal force: $\left.\hat{n}_\pm \cdot \bm{\sigma}\cdot\hat{n}_\pm\right|_{x=\pm L}=-\gamma  \vec{\nabla}\cdot\hat{n}_\pm|_{x=\pm L}\approx \pm \gamma\, \dd^2\delta L/\dd y^2$. Then, the expansion of the growth rate at long wavelengths reads
\begin{equation}
\omega\left(q\right)\approx \frac{1}{\tau} + \left[\frac{1}{3\tau} - \frac{\gamma}{2\eta L_0}\right] \left(qL_0\right)^2 + \mathcal{O}\left(\left(qL_0\right)^4\right),
\end{equation}
with $\tau=2\eta h/(T_0L_c)$. This expression reveals the existence of a critical size $L_\gamma\approx 3\gamma h/(T_0 L_c)$ above which the growth rate curves upwards at $q\rightarrow 0$ ($\omega''(0)>0$), hence exhibiting the aforementioned peak at a finite wavelength. Alternatively, if $\gamma>\gamma^* \approx T_0 L_c L_0/(3h)$, surface tension prevents the hydrodynamic selection of a finite fingering wavelength, which is then only limited by the length of the tissue front.

The surface tension of the monolayer could be due to actin cables found along its edge, specially along the sides of epithelial fingers \cite{Poujade2007,Klarlund2012,Reffay2014}. Traction force measurements suggest that the tension of such cables is $\gamma\sim 0.2$ mN/m \cite{Reffay2014}, lower than typical surface tensions of cell aggregates, $\gamma\sim 1-10$ mN/m \cite{Foty1994,Forgacs1998,Guevorkian2010,Stirbat2013}. Combining these value ranges with $T_0=0.2-0.8$ kPa and typical values of $h$ and $L_c$ (\cref{t parameters}), the critical monolayer width for fingering is $L_\gamma\sim 0.3-10$ $\mu$m. Therefore, we expect surface tension not to play a major role in the fingering instability in monolayers of typical widths $L_0\sim 0.1-1$ mm.

\paragraph{Intercellular contractility.---} Because it decreases the spreading velocity, intercellular contractility has an additional stabilizing effect on the monolayer front (\cref{fig origins}d). We discuss the effects of a uniform contractility in \cite{SM}. Here, we consider an intercellular contractility $-\zeta$ concentrated at the polarized boundary layer, which has a size-independent contribution to the spreading velocity:
\begin{equation}
V_0\approx \frac{T_0 L_c}{2\eta h}L_0 + \frac{L_c}{2\eta}\left[\frac{\zeta}{2}-\frac{T_0 L_c}{h}\right].
\end{equation}
Consequently, in the limit $L_c\ll L_0\ll \lambda$, this contractility has no impact on the growth rate of the uniform mode, $\omega\left(0\right)=\tau^{-1}$, but it contributes a stabilizing quadratic term to the long-wavelength expansion of the growth rate:
\begin{equation}
\omega\left(q\right)\approx \frac{1}{\tau} + \left[\frac{1}{3\tau} + \frac{\zeta L_c^2}{8\eta L_0^2}\right] \left(qL_0\right)^2 + \mathcal{O}\left(\left(qL_0\right)^4\right).
\end{equation}
Thus, as surface tension, contractility defines a critical size $L_\zeta\approx \sqrt{-3\zeta L_c h/(4T_0)}$ above which the growth rate features a fi\-nite-wave\-length peak. Alternatively, if $-\zeta>-\zeta^*\approx 4 T_0 L_0^2/(3 h L_c)$, contractility supresses the hydrodynamic selection of a finite fingering wavelength.

For typical contractilities, $-\zeta\sim 1-50$ kPa \cite{Perez-Gonzalez2019}, the critical monolayer width for fingering is $L_\zeta\sim 10-100$ $\mu$m, where we have used $T_0=0.2-0.8$ kPa and estimates for $h$ and $L_c$ (\cref{t parameters}). Therefore, we do not expect contractility to prevent fingering wavelength selection. However, our estimates give $L_\gamma<L_\zeta$, indicating that contractility typically dominates over surface tension in stabilizing short-wave\-length shape perturbations. Thus, the competition between the destabilizing effect of traction forces and the stabilizing effect of intercellular contractility defines the band of unstable modes.

\paragraph{Uniformly polarized tissues.---} To consider monolayers with bulk polarity \cite{Trepat2009}, we analyze the morphological stability of a uniformly polarized monolayer \cite{SM}. In this case, in the absence of contractility, the spreading velocity is constant: $V_0= T_0/(\xi h)$. Hence, the $q=0$ mode is marginally stable, $\omega(q=0)=0$. Moreover, contractility stabilizes it, $\omega(0)<0$. Nevertheless, the viscous effects discussed above still give rise to a peak of the growth rate at a finite wavelength. Therefore, even though uniformly polarized tissues do not feature an accelerating front, they still exhibit a fingering instability for sufficiently small contractility \cite{SM}.

\paragraph{Conclusions.---} Motivated by the observation of finger-like protrusions during the spreading of epithelial monolayers, we studied the stability of the advancing front. Modeling the cell monolayer as an active polar fluid, we showed that active traction forces are responsible for a long-wave\-length instability of the monolayer front. Several features distinguish this instability from previous proposals. First, it is generic; it takes place for any value of the active traction force. Second, the wavelength of the fingering pattern is selected by the range of hydrodynamic interactions in the tissue. And third, active intercellular forces stabilize short-wave\-length perturbations, typically dominating over surface tension effects.

Our analysis identifies the physical mechanism of the instability. Cellular traction forces at the monolayer edge set the velocity gradient in the spreading monolayer. Hence, under the same traction force, a larger monolayer spreads faster \cite{Perez-Gonzalez2019}. Consequently, when the monolayer front is perturbed, the protruding regions of the interface advance faster than the trailing regions, thus making the perturbation grow. Therefore, the instability is based on a simple kinematic mechanism, which takes place generically in viscous fluids that sustain a fixed velocity gradient in the direction of spreading. In particular, the same morphological instability should occur in the so-called squeeze flow \cite{Engmann2005}, in which an incompressible fluid is forced to spread by decreasing the gap between two plates. In this case, under perfect slip conditions at the plates, the rate of gap reduction sets the fixed velocity gradient.

Regarding spreading epithelia, we conclude that neither leader-cell behavior nor regulation of cell motility by curvature or by chemotactic fields are necessary for the fingering instability. Therefore, our results are consistent with the emergence of leader cells concomitantly with finger growth \cite{Poujade2007,Petitjean2010,Reffay2011,Reffay2014,Vishwakarma2018}. However, the viscous rheology of the monolayer is essential for the instability. On the one hand, it underpins the velocity gradient that renders the interface unstable and, on the other, it enables wavelength selection for the fingering pattern. Concomitant with the fingering instability, shear stresses give rise to flows transversal to the spreading direction, which might lead to the swirls observed in experiments \cite{Poujade2007,Petitjean2010}. Finally, in addition to explaining fingering in tissue spreading, our results also account for the morphological instability recently observed during tissue dewetting \cite{Perez-Gonzalez2019}.

Our predictions, such as the absence of a traction force threshold for the instability, and whether the fingering wavelength is given by either the monolayer width or the screening length $\lambda=\sqrt{\eta/\xi}$, are experimentally testable. Indeed, consistent with our result, recent work has shown that the finger spacing is an intrinsic quantity that coincides with the stress correlation length \cite{Vishwakarma2018}. To further test our predictions, future experiments could perturb active cellular forces, cell-cell and cell-substrate adhesion, and vary monolayer width.

Our findings illustrate how hydrodynamic interactions impact tissue morphodynamics. In particular, we propose that epithelial fingering can naturally arise from a generic morphological instability in a fluid film driven by interfacial active forces. Thus, our results showcase the relevance of interfacial instabilities in driven \cite{Troian1989,Melo1989,BenAmar2001} and active \cite{Callan-Jones2008,Sankararaman2009,Sarkar2012,Sarkar2013a,Nagilla2018,Williamson2018,Bogdan2018} fluids for tissue spreading.

\begin{acknowledgements}
We thank Xavier Trepat for a critical reading of the manuscript, and the members of his lab for discussions. We thank Jordi Ort\'{i}n for discussions. R.A. acknowledges support from Fundaci\'{o} ``La Caixa''. R.A. and J.C. acknowledge the MINECO under project FIS2016-78507-C2-2-P and Generalitat de Catalunya under project 2014-SGR-878.
\end{acknowledgements}

\bibliography{All}


\onecolumngrid 

\clearpage
\appendix

\onecolumngrid
\begin{center}
\textbf{\large Supplementary Material for ``Active Fingering Instability in Tissue Spreading''}
\end{center}

\setcounter{equation}{0}
\setcounter{figure}{0}
\renewcommand{\theequation}{S\arabic{equation}}
\renewcommand{\thefigure}{S\arabic{figure}}

\twocolumngrid

\section{ACTIVE POLAR FLUID MODEL OF EPITHELIAL SPREADING} \label{active-polar-fluid-model}

Here, we briefly justify the description of epithelial spreading in terms of our continuum active polar fluid model.

\subsection{Polarity dynamics}

The outwards polarization of cells at the monolayer edge is likely due to contact inhibition of locomotion, a cell-cell interaction whereby cells repolarize in opposite directions upon contact \cite{Mayor2010,Stramer2017}. In fact, this interaction is mediated by cell-cell adhesion, with front-rear differences in cadherin-based junctions acting as a cue for the repolarization \cite{Desai2009,Khalil2010,Weber2012,Theveneau2013,Vedula2013a,Ladoux2016}. Although originally proposed for mesenchymal cells, contact inhibition of locomotion is being increasingly recognized to play a key role in orchestrating the collective migration of epithelial monolayers \cite{Mayor2010,Theveneau2013,Vedula2013a,Ladoux2016,Mayor2016,Hakim2017a,Zimmermann2016a,Coburn2016,Smeets2016}. In a cohesive monolayer, this interaction naturally leads to polarization of cells at the edge towards free space, leaving the inner region of the monolayer unpolarized. Such a polarity profile, in turn, explains the localization of traction forces at the edge and the build-up of tension at the center of epithelial monolayers \cite{Zimmermann2016a,Coburn2016}. Therefore, we assume the polarity field $\vec{p}\left(\vec{r},t\right)$ to be set by an autonomous cellular mechanism such as contact inhibition of locomotion, which polarizes cells within a time scale $\tau_{\text{CIL}}\sim 10$ min \cite{Smeets2016,Weber2012}. Hence, $\vec{p}\left(\vec{r},t\right)$ should remain essentially independent of flows in the monolayer, which occur over a longer time scale given by the strain rate, at least of order $\tau_s\sim 100$ min \cite{Blanch-Mercader2017,Vincent2015}. Consequently, taking a phenomenological approach, we propose the polarity field to follow a purely relaxational dynamics given by
\begin{equation} \label{eq polarity-dynamics}
\frac{\partial p_\alpha}{\partial t}=-\frac{1}{\gamma_1}\frac{\delta F}{\delta p_\alpha},
\end{equation}
where $F\left[\vec{p}\,\right]$ is the coarse-grained free energy functional for the polarity field (\cref{eq polar-free-energy}), and $\gamma_1$ is a kinetic coefficient (the rotational viscosity for the angular degree of freedom). With respect to the most general dynamics of the polarity field in an active polar fluid, \cref{eq polarity-dynamics} neglects polarity advection and corotation, as well as flow alignment and active spontaneous polarization effects. Thus, using \cref{eq polar-free-energy}, the dynamics of the polarity is given by
\begin{equation}
\partial_t p_\alpha=\frac{1}{\gamma_1}\left(-a p_\alpha + K\nabla^2 p_\alpha\right).
\end{equation}
In the limit of fast polarity dynamics compared to the spreading dynamics, the polarity field is always at equilibrium, $\partial_t p_\alpha=0$, adiabatically adapting to the shape of the monolayer. Under this approximation, the polarity field is given by
\begin{equation} \label{eq polarity-field-monolayer}
L_c^2 \nabla^2 p_\alpha= p_\alpha,
\end{equation}
where we have defined the characteristic length $L_c\equiv \sqrt{K/a}$ of the polar order in the monolayer.

\subsection{Force balance}

Flows in cell monolayers occur at very low Reynolds numbers. Therefore, inertial forces are negligible, and hence momentum conservation reduces to the force balance condition
\begin{equation}
0 = \partial_\beta \sigma_{\alpha\beta} + f_\alpha,\qquad \sigma_{\alpha\beta}=\sigma_{\alpha\beta}^s+\sigma_{\alpha\beta}^a+\sigma_{\alpha\beta}^{E,s}
\end{equation}
where $\sigma_{\alpha\beta}^s$ and $\sigma_{\alpha\beta}^a$ are the symmetric and antisymmetric parts of the deviatoric stress tensor, and $f_\alpha$ is the external force density. Respectively, $\sigma_{\alpha\beta}^{E,s}$ is the symmetric part of the Ericksen tensor. This tensor generalizes the pressure $P$ to include anisotropic elastic stresses associated to the orientational degrees of freedom in liquid crystals \cite{deGennes-Prost}:
\begin{equation} \label{eq Ericksen-tensor}
\sigma_{\alpha\beta}^{E}=-P\,\delta_{\alpha\beta} - \frac{\partial f}{\partial\left(\partial_\beta p_\gamma\right)}\partial_\alpha p_\gamma,
\end{equation}
where $f$ is the Frank free energy density, namely the integrand of \cref{eq polar-free-energy}. Thus, the orientational contribution to the Ericksen tensor is of second order in gradients of the polarity field, and hence we neglect it, so that the stress tensor reads
\begin{equation}
\sigma_{\alpha\beta}=\sigma_{\alpha\beta}^s+\sigma_{\alpha\beta}^a - P\,\delta_{\alpha\beta}.
\end{equation}

Then, the pressure is related to the cell number surface density $\rho$ by the equation of state of the monolayer. For the sake of an estimate, we assume the simplest form for an equation of state, $P\left(\rho\right)=B \left(\rho-\rho_0\right)/\rho_0$, where $B$ is the bulk modulus of the monolayer, and $\rho_0$ is a reference density defined by $P\left(\rho_0\right)=0$. Taking the pressure origin at the monolayer edge, $\rho_0\sim 2\cdot 10^3$ cells/mm$^2$ \cite{Trepat2009,Perez-Gonzalez2019}. In turn, density differences in the monolayer are, at most, $\rho-\rho_0\sim 6\cdot 10^3$ cells/mm$^2$ \cite{Trepat2009,Perez-Gonzalez2019}. Then, the monolayer is expected to be highly compressible because area changes can in principle be accommodated by changes in height, resisted only by the shear modulus of the tissue. Hence, we estimate the bulk modulus of the monolayer by typical shear moduli of cell aggregates, which are in the range $G\sim 10^2-10^3$ Pa \cite{Forgacs1998,Marmottant2009,Guevorkian2010}. Thus, the pressure in the monolayer should be $P\lesssim 30-300$ Pa. In fact, isotropic compressive stresses (pressures) of $\sim 50$ Pa were shown to induce cell extrusion \cite{Saw2017}. In conclusion, if tissue spreading is not dominated by cell proliferation \cite{Basan2013,Recho2016,Yabunaka2017a}, the magnitude of the pressure in the monolayer is expected to be much smaller than the tensile stress (tension) induced by traction forces, as measured by monolayer stress microscopy, which is of the order of kPa \cite{Trepat2009,Perez-Gonzalez2019}. Hence, we neglect pressure:
\begin{equation}
\sigma_{\alpha\beta}=\sigma_{\alpha\beta}^s+\sigma_{\alpha\beta}^a.
\end{equation}

Now, for a nematic medium, the antisymmetric part of the stress tensor is given by $\sigma_{\alpha\beta}^a=1/2\left(p_\alpha h_\beta - h_\alpha p_\beta\right)$, where $h_\alpha=-\delta F/\delta p_\alpha$ is the molecular field. From \cref{eq polarity-dynamics}, the adiabatic approximation for the polarity dynamics, $\partial_t p_\alpha=0$, implies $h_\alpha=0$. Therefore, the antisymmetric part of the stress tensor vanishes under this approximation, $\sigma_{\alpha\beta}^a=0$. Thus, the stress tensor reduces to
\begin{equation} \label{eq force-balance-monolayer}
\sigma_{\alpha\beta}= \sigma_{\alpha\beta}^s.
\end{equation}

\subsection{Constitutive equations}

Next, constitutive equations must be given to specify the deviatoric stress tensor $\sigma_{\alpha\beta}^s$ and the external force $f_\alpha$ in terms of the polarity and velocity fields. The generic constitutive equations of an active liquid crystal are provided by active gel theory \cite{Kruse2005,Julicher2011,Marchetti2013,Prost2015}. Here, based on the previous assumptions for the dynamics of the polarity field, we propose a simplified version of the generic constitutive equations of an active polar gel to describe epithelial spreading.

First, the spreading occurs on timescales of the order of $\tau_s\sim 100$ min \cite{Blanch-Mercader2017,Vincent2015}, at which the tissue should have a fluid rheology. This time scale is much slower than the turnover time scales of proteins in the cytoskeleton or in cell-cell junctions, which are of the order of tens of minutes at most \cite{Wyatt2016,Khalilgharibi2016}. Intra- or intercellular processes such as cytoskeletal reorganizations or cell-cell slidings dissipate energy over these time scales, so that elastic energy may only be stored in the tissue at shorter times. In addition, other processes such as cell division, death, and extrusion \cite{Ranft2010,Matoz-Fernandez2017a}, as well as cell shape fluctuations \cite{Marmottant2009,Etournay2015} and topological rearrengements \cite{Etournay2015,Krajnc2018a} also fluidize the tissue. Therefore, to describe the slow spreading dynamics, we will not consider the elastic response of the tissue at short time scales.

Then, in the viscous limit, the constitutive equations for the internal stress and the interfacial force of an active polar medium are:
\begin{multline} \label{eq constitutive-equation-stress}
\sigma_{\alpha\beta}^s=2\eta \tilde{v}_{\alpha\beta}+\frac{\nu_1}{2}\left(p_\alpha h_\beta + h_\alpha p_\beta -\frac{2}{d} p_\gamma h_\gamma \delta_{\alpha\beta}\right) - \zeta q_{\alpha\beta}\\
+ \left(\bar{\eta}\, d\, v_{\gamma\gamma} + \bar{\nu}_1\, d\, p_\gamma h_\gamma - \bar{\zeta} - \zeta' p_\gamma p_\gamma\right) \delta_{\alpha\beta},
\end{multline}
\begin{equation} \label{eq constitutive-equation-interfacial-force}
f_\alpha = - \xi v_\alpha + \nu_i \dot{p}_\alpha + \zeta_i p_\alpha,
\end{equation}
where, $q_{\alpha\beta}=p_\alpha p_\beta - p_\gamma p_\gamma/d\; \delta_{\alpha\beta}$ is the traceless symmetric nematic order parameter tensor, with $d$ the system dimensionality, and $v_\alpha$ is the velocity of the fluid with respect to the substrate. The coefficients $\eta$ and $\bar{\eta}$ are the shear and bulk viscosities of the medium, $\zeta$ is the anisotropic active stress coefficient, and $\bar{\zeta}$ and $\zeta'$ are two isotropic active stress coefficients. Finally, $\xi$, $\nu_i$, and $\zeta_i$ are the corresponding interfacial versions of the viscosity (viscous friction), flow alignment (polar friction), and active stress (active force) coefficients. The constitutive equation for the internal stress, \cref{eq constitutive-equation-stress}, is that of an active polar gel with a variable modulus of the polarity \cite{Julicher2011}. In turn, the constitutive equation for the interfacial force, \cref{eq constitutive-equation-interfacial-force}, is less conventional \cite{Julicher2009}, but it was derived from a mesoscopic model of an active polar gel \cite{Oriola2017}.

Now, the adiabatic approximation for the polarity dynamics implies $\dot{p}_\alpha=h_\alpha=0$, so that flow alignment terms contribute neither to the stress tensor nor to the interfacial force. Next, we assume that polarized cells generate much larger active stresses than unpolarized cells. Hence, we neglect the active stress coefficient $\bar{\zeta}$ in front of $\zeta$ and $\zeta'$. Then, assuming that the two-dimensional fluid layer is compressible, we take $\zeta=\zeta'\,d=2\zeta'$ and $2\eta=\bar{\eta}\, d=2\bar{\eta}$ for simplicity. Under these simplifications, and using \cref{eq force-balance-monolayer}, the constitutive equations reduce to
\begin{equation} \label{eq bulk-constitutive-equation-monolayer}
\sigma_{\alpha\beta}=\eta \left(\partial_\alpha v_\beta + \partial_\beta v_\alpha\right) - \zeta p_\alpha p_\beta,
\end{equation}
\begin{equation} \label{eq interfacial-constitutive-equation-monolayer}
f_\alpha= - \xi v_\alpha + \zeta_i p_\alpha,
\end{equation}
which close the set of equations defining the active polar fluid model of the spreading of an epithelial monolayer.

\section{LINEAR STABILITY ANALYSIS} \label{LSA}

Here, we give the details of the linear stability analysis of the tissue front. First, we explicitly write down the equations of the model in Cartesian coordiates, which are most convenient for the rectangular geometry of the monolayer (\cref{fig instability}a). Thus, the equation for the polarity field, \cref{eq polarity-field}, reads
\begin{subequations} \label{eq polarity-components-instability}
\begin{align}
L_c^2\left(\partial_x^2 + \partial_y^2\right)p_x&=p_x,\\
L_c^2\left(\partial_x^2 + \partial_y^2\right)p_y&=p_y.
\end{align}
\end{subequations}
Respectively, the force balance equation \cref{eq force-balance} reads
\begin{subequations} \label{eq force-balance-components-instability}
\begin{align}
\partial_x \sigma_{xx} +\partial_y \sigma_{xy}&=\xi v_x - T_0/h \, p_x,\\
\partial_x \sigma_{yx} +\partial_y \sigma_{yy}&=\xi v_y - T_0/h \, p_y,
\end{align}
\end{subequations}
where the components of the stress tensor are given by
\begin{subequations} \label{eq stress-tensor-components-instability}
\begin{align}
\sigma_{xx}&=2\eta\, \partial_x v_x - \zeta p_x^2,\\
\sigma_{xy}&=\sigma_{yx}=\eta\left(\partial_x v_y+\partial_y v_x\right)-\zeta p_x p_y,\\
\sigma_{yy}&=2\eta\, \partial_y v_y - \zeta p_y^2.
\end{align}
\end{subequations}

Next, we obtain the flat front solution in rectangular geometry, which is the reference state of the linear stability analysis. The long ($\hat{y}$) axis of the rectangle is much longer than the short ($\hat{x}$) axis. Hence, we assume translational invariance along the long axis of the monolayer \cite{Blanch-Mercader2017}. Moreover, traction forces are mainly perpendicular to the monolayer boundary \cite{Trepat2009}, so that we take the polarity field along the $\hat{x}$ direction: $\vec{p}=p_x^0\left(x\right)\hat{x}$, where the superindex indicates the zeroth order in the perturbations of the front. Now, imposing symmetry as well as maximal polarity and stress-free boundary conditions, $p_x^0\left(L_0\right)=1$ and $\sigma_{xx}^0 \left(L_0\right)=0$, one obtains the polarity and velocity profiles:
\begin{equation}
p_x^0\left(x\right)=\frac{\sinh\left(x/L_c\right)}{\sinh\left(L_0/L_c\right)},
\end{equation}
\begin{multline}
v_x^0\left(x\right) = \frac{\bar{\lambda}}{2\eta}\left[\zeta+\frac{T_0L_c\bar{\lambda}^2/h}{\bar{\lambda}^2-L_c^2}\coth\left(L_0/L_c\right) \right.\\
\left. - \frac{2\zeta \bar{\lambda}^2}{4\bar{\lambda}^2-L_c^2}\left[2+\csch^2\left(L_0/L_c\right)\right]\right]\frac{\sinh\left(x/\bar{\lambda}\right)}{\cosh\left(L_0/\bar{\lambda}\right)}\\
+\frac{L_c}{\xi\sinh\left(L_0/L_c\right)}\\
\left[\frac{\zeta}{4\bar{\lambda}^2-L_c^2}\frac{\sinh\left(2x/L_c\right)}{\sinh\left(L_0/L_c\right)}-\frac{T_0L_c/h}{\bar{\lambda}^2-L_c^2}\sinh\left(x/L_c\right)\right],
\end{multline}
where $\bar{\lambda}=\sqrt{2\eta/\xi}=\sqrt{2}\,\lambda$ is a redefined hydrodynamic screening length, and $L_0$ is the semi-width of the monolayer, which changes according to $\dd L_0/\dd t=v_x^0\left(L_0\right)$.

Next, we introduce peristaltic small-amplitude perturbations of the flat interface of the monolayer (\cref{fig instability}a):
\begin{equation}
L\left(y\right)=L_0 + \delta L\left(y\right).
\end{equation}
Under these perturbations, the polarity and velocity fields take the form
\begin{align}
p_x\left(x,y\right)=p_x^0\left(x\right)+\delta p_x\left(x,y\right),\quad p_y\left(x,y\right)=\delta p_y\left(x,y\right), \label{eq pert-polarity}\\
v_x\left(x,y\right)=v_x^0\left(x\right)+\delta v_x\left(x,y\right),\quad v_y\left(x,y\right)=\delta v_y\left(x,y\right). \label{eq pert-velocity}
\end{align}
In turn, boundary conditions must keep imposing a normal and maximal polarity, as well as vanishing normal and shear stresses at the interface, which is now curved. To this end, we define the normal and tangential vectors of each interface,
\begin{subequations}
\begin{align}
\hat{n}_\pm & = \pm \cos\theta \, \hat{x} + \sin\theta \, \hat{y} \approx \pm \, \hat{x} - \frac{\dd\delta L}{\dd y} \, \hat{y},\\
\hat{t}_\pm & = \mp \sin\theta \, \hat{x} + \cos\theta \, \hat{y} \approx \pm \frac{\dd\delta L}{\dd y} \, \hat{x} + \hat{y},
\end{align}
\end{subequations}
where $\theta$ is the angle between the normal directions of the flat and perturbed interfaces, and the $\pm$ index stands for the top and bottom interfaces, respectively (\cref{fig instability}a). Thus, the boundary conditions for the polarity read
\begin{equation}
\left.\vec{p}\cdot\hat{n}_\pm\right|_{x=\pm L}=1,\qquad \left.\vec{p}\cdot\hat{t}_\pm\right|_{x=\pm L}=0.
\end{equation}
For the $x$-component, the conditions imply $p_x\left(\pm L\right)\approx \pm 1$. This expands into
\begin{multline}
p_x\left(\pm L\right)=p_x^0\left(\pm L\right)+\delta p_x\left(\pm L\right)\\
\approx p_x^0\left(\pm L_0\right)\pm \partial_x p_x^0\left(\pm L_0\right)\delta L+\delta p_x\left(\pm L\right)\approx \pm1,
\end{multline}
which yields
\begin{equation}
\delta p_x\left(\pm L\right)=\mp \partial_x p_x^0\left(\pm L_0\right)\delta L
\end{equation}
as a boundary condition on the polarity perturbation. For the $y$-component of the polarity perturbation, the boundary condition imposes
\begin{equation}
\delta p_y\left(\pm L\right)=-\frac{\dd\delta L}{\dd y}.
\end{equation}
Then, the boundary conditions on the stress read
\begin{equation} \label{eq boundary-conditions-stress-instability}
\left.\hat{n}_\pm\cdot\bm{\sigma}\cdot \hat{n}_\pm\right|_{x=\pm L}=0,\qquad \left.\hat{t}_\pm\cdot\bm{\sigma}\cdot \hat{n}_\pm\right|_{x=\pm L}=0,
\end{equation}
which respectively ensure vanishing normal and shear stress at the interfaces. Here, for simplicity, we neglect interfacial tension and bending rigidity, which would contribute stabilizing terms to the growth rate. The condition on the normal stress gives $\sigma_{xx}\left(\pm L\right)=0$ which, after expanding as previously, leads to
\begin{equation}
\delta\sigma_{xx}\left(\pm L\right)=\mp \partial_x\sigma_{xx}^0\left(\pm L_0\right)\delta L
\end{equation}
for the stress perturbation. In turn, the condition on the shear stress directly gives
\begin{equation}
\delta\sigma_{xy}\left(\pm L\right)=0.
\end{equation}

Next, we decompose all perturbations in their Fourier modes, identified by the wave number $q$:
\begin{subequations}
\begin{align}
\delta L\left(y,t\right)&=\int_{-\infty}^\infty \delta \tilde{L}\left(q,t\right) e^{iqy}\, \frac{\dd q}{2\pi},\\
\delta p_\alpha\left(x,y,t\right)&=\int_{-\infty}^\infty \delta \tilde{p}_\alpha\left(x,q,t\right) e^{iqy} \,\frac{\dd q}{2\pi},\\
\delta v_\alpha\left(x,y,t\right)&=\int_{-\infty}^\infty \delta \tilde{v}_\alpha\left(x,q,t\right) e^{iqy} \,\frac{\dd q}{2\pi},
\end{align}
\end{subequations}
In terms of the Fourier modes, the equations for the polarity components read
\begin{subequations} \label{eq polarity-Fourier-instability}
\begin{align}
L_c^2\left(\partial_x^2-q^2\right)\delta\tilde{p}_x=\delta\tilde{p}_x,\\
L_c^2\left(\partial_x^2-q^2\right)\delta\tilde{p}_y=\delta\tilde{p}_y.
\end{align}
\end{subequations}
In turn, the components of the force balance equation, once the constitutive relation is introduced, read
\begin{subequations} \label{eq velocity-Fourier-instability}
\begin{gather}
\begin{multlined}
\eta\left(2\partial_x^2-q^2-\frac{1}{\lambda^2}\right)\delta\tilde{v}_x+iq\eta\,\partial_x \delta\tilde{v}_y\\
+\left[T_0/h-2\zeta\left(\partial_x p_x^0+p_x^0\partial_x\right)\right]\delta\tilde{p}_x-iq\zeta\, p_x^0\, \delta\tilde{p}_y=0,
\end{multlined}
\\
\begin{multlined}
iq\eta\,\partial_x\delta\tilde{v}_x+\eta\left(\partial_x^2-2q^2-\frac{1}{\lambda^2}\right)\delta\tilde{v}_y\\
+\left[T_0/h-\zeta\left(\partial_x p_x^0+p_x^0\partial_x\right)\right]\delta\tilde{p}_y=0.
\end{multlined}
\end{gather}
\end{subequations}
The boundary conditions must also be translated into the Fourier domain, reading
\begin{equation} \label{eq boundary-conditions-Fourier-instability}
\delta\tilde{p}_x\left(\pm L\right)=\mp \partial_x p_x^0\left(\pm L_0\right)\delta\tilde{L},\qquad \delta\tilde{p}_y\left(\pm L\right)=-iq\,\delta\tilde{L},
\end{equation}
\begin{equation} \label{eq boundary-conditions-stress-Fourier-instability}
\delta\tilde{\sigma}_{xx}\left(\pm L\right)=\mp \partial_x\sigma_{xx}^0\left(\pm L_0\right)\delta\tilde{L},\qquad \delta\tilde{\sigma}_{xy}\left(\pm L\right)=0.
\end{equation}

Then, the four coupled differential equations \cref{eq polarity-Fourier-instability,eq velocity-Fourier-instability} are analytically solved for $\delta\tilde{p}_\alpha \left(x,q\right)$ and $\delta\tilde{v}_\alpha \left(x,q\right)$. From the Fourier modes of the velocity field, the perturbed spreading velocity $V$ can be computed as
\begin{multline}
V=\left.\vec{v}\cdot\hat{n}\right|_{x=L}=\left[\vec{v}\,^0 \cdot \hat{n} + \delta\vec{v}\cdot\hat{n}\right]_{x=L}\\
\approx v_x^0\left(L_0\right) + \partial_x v_x^0\left(L_0\right) \delta L + \delta v_x\left(L_0\right),
\end{multline}
so that
\begin{equation}
\delta V\left(y\right)=V\left(y\right)-V_0=\partial_x v_x^0\left(L_0\right) \delta L\left(y\right) + \delta v_x\left(L_0,y\right).
\end{equation}
Thus, the growth rate $\omega\left(q\right)$ of the tissue shape perturbations follows from
\begin{equation} \label{eq growth-rate-definition}
\delta\tilde{V}\left(q\right)=\int_{-\infty}^\infty \delta V\left(y\right) e^{-iqy} dy = \frac{\dd\delta\tilde{L}\left(q\right)}{\dd t}=\omega\left(q\right) \delta\tilde{L}\left(q\right).
\end{equation}
Hence,
\begin{equation} \label{eq growth-rate}
\omega\left(q\right)=\partial_x v_x^0\left(L_0\right) + \frac{\delta \tilde{v}_x\left(L_0,q\right)}{\delta\tilde{L}\left(q\right)}.
\end{equation}
The expression of the resulting growth rate is omitted here due to its length. Finally, note that, in our free-boundary problem, the amplitude of the front perturbations does not grow exponentially in time. This is because the growth rate depends on time through the monolayer width $L_0\left(t\right)$. Consequently, \cref{eq growth-rate-definition} yields
\begin{equation}
\delta\tilde{L}\left(q,t\right)=\delta\tilde{L}\left(q,0\right)\exp\left[\int_0^t \omega\left(q,t'\right) \dd t'\right].
\end{equation}

\section{NEMATIC ELASTICITY}

In this section, we discuss the effects of the nematic elasticity of the polarity field on the growth rate of front shape perturbations. Front perturbations distort the polarity field, generating polarity gradients along the tissue front. The elastic energy cost of these transversal polarity gradients is larger for shorter-wavelength perturbations. Thus, to minimize the polar free energy \cref{eq polar-free-energy}, polarity perturbations decay more steeply for shorter wavelengths. From \cref{eq polarity-field}, the Fourier components of polarity perturbations obey $L_c^2 \left(\partial_x^2 - q^2\right)\delta\tilde{p}_\alpha=\delta \tilde{p}_\alpha$, so that their decay length is $\ell_c(q)=L_c[1+(q L_c)^2]^{-1/2}$, which decreases with $q$. Similarly to \cref{eq front-acceleration}, the corresponding velocity gradient perturbation is then proportional to $T_0\ell_c(q)$. Therefore, by reducing the size of the polarized boundary layer, $\ell_c(q)\leq L_c$, nematic elasticity causes a decrease of the growth rate with decreasing wavelength. However, this effect is only significant at wavelengths shorter than the nematic length $L_c$ ($qL_c>1$), which is typically smaller than the size of the fingers.

\section{UNIFORM INTERCELLULAR CONTRACTILITY}

\begin{figure}[tb]
\begin{center}
\includegraphics[width=0.7\columnwidth]{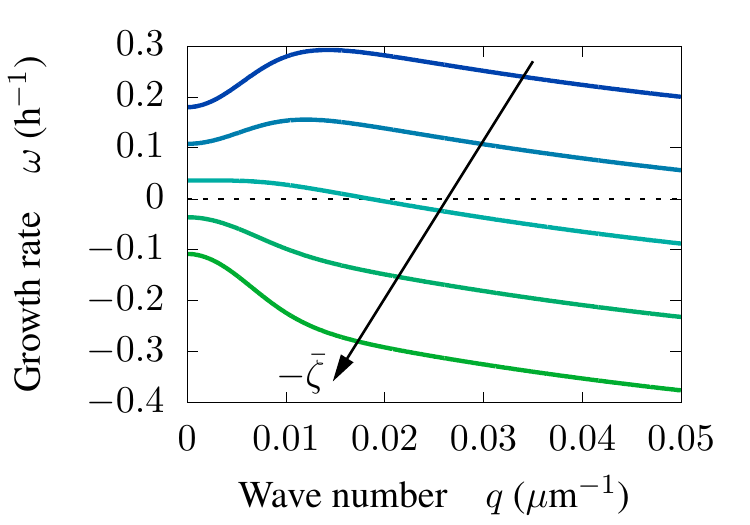}
\caption{Growth rate of shape perturbations for different values of the uniform intercellular contractility $-\bar{\zeta}=0,-1,-2,-3,-4$ kPa, which has a stabilizing effect on perturbations of all wavelengths. The other parameter values are in Table I in the Main Text except for $\xi,\zeta\rightarrow 0$.} \label{fig uniform-contractility}
\end{center}
\end{figure}

In this section, we discuss the effects of a uniform intercellular contractility term, characterized by the coefficient $-\bar{\zeta}$ in \cref{eq constitutive-equation-stress}, which is neglected in the Main Text. Like the intercellular contractility $-\zeta$, which is localized at the polarized boundary layer of the tissue, a uniform contractility has a stabilizing effect on the tissue front (\cref{fig uniform-contractility}). However, unlike its polarity-related counterpart $-\zeta$, the uniform contractility $-\bar{\zeta}$ has a size-dependent contribution to the spreading velocity:
\begin{equation}
V_0\approx \frac{1}{2\eta}\left[\frac{T_0L_c}{h}+\bar{\zeta}\right]L_0
\end{equation}
in the limit $L_c\ll L_0\ll \lambda$. Then, the expansion of the growth rate at long wavelengths reads
\begin{equation}
\omega\left(q\right)\approx \frac{1}{\tau} + \frac{\bar{\zeta}}{2\eta} + \left[\frac{1}{3\tau} + \frac{\bar{\zeta}}{3\eta}\right] \left(qL_0\right)^2 + \mathcal{O}\left(\left(qL_0\right)^4\right),
\end{equation}
where $\tau=2\eta h/(T_0L_c)$. This expression shows that a uniform contractility does not only affect the growth rate of finite-wavelength perturbations but that it also decreases the growth rate of the uniform mode, $\omega(0)$. If the contractility is sufficiently small to allow the tissue to spread ($V_0>0$), $-\bar{\zeta}<T_0L_c/h$, the tissue front remains unstable to long-wavelength perturbations. However, unlike the polarity-related contractility $-\zeta$, if the uniform contractility $-\bar{\zeta}$ induces the retraction of the monolayer front, it also prevents its fingering instability (\cref{fig uniform-contractility}).

\section{LINEAR STABILITY ANALYSIS FOR UNIFORMLY POLARIZED TISSUES} \label{polarized}

In this section, we address situations in which the epithelial monolayer is polarized not only in a boundary layer but also in its bulk \cite{Trepat2009}. To analyze the effects of bulk polarity, we consider the simplest possible situation, namely a uniformly polarized epithelium. The state of uniform polarization, however, presents two issues. First, it may be unstable in the bulk \cite{Blanch-Mercader2017c}. Here, we assume that the active contractility and traction force coefficients are small enough not to trigger this bulk instability. Second, because the polarity points outward at both fronts of the monolayer (\cref{fig instability}), a spreading monolayer cannot have a uniform polarity. Following previous works \cite{Zimmermann2014a,Nesbitt2017a}, we avoid this issue by considering the tissue to be bounded by a comoving wall at $x=0$. In this case, we can assume a fixed modulus of the polarity field, $|\vec{p}\,|=1$. Hence, the polar free energy is left only with Frank elasticity:
\begin{equation}
F = \int \frac{K}{2} (\partial_\alpha p_\beta)(\partial_\alpha p_\beta)\,\dd^3\vec{r}.
\end{equation}
Consequently, the equilibrated polarity field, which fulfills $\delta F/\delta p_\alpha = 0$, is a solution of
\begin{equation}
\nabla^2 p_\alpha = 0.
\end{equation}
The rest of the model is unchanged with respect to the case with only a polarized boundary layer analyzed in the Main Text. Thus, the force balance is still given by Eqs. 3-4, and the boundary conditions still impose a normal polarity and a vanishing stress at the monolayer front: $\vec{p}(x=L)=\hat{n}$ and $\left.\bm{\sigma}\cdot\hat{n}\right|_{x=L} = 0$, respectively, with $\hat{n}$ being the normal unit vector of the tissue front. Moreover, we impose the following boundary conditions at the comoving wall: $\vec{p}(x=0)=\hat{x}$, and $\vec{v}(x=0)= T_0/(\xi h)\,\hat{x}$. Thus, the reference state of the stability analysis is the flat front solution with
\begin{equation}
p_x^0=1,\qquad v_x^0(x) = \frac{1}{\xi}\left[\frac{T_0}{h} + \frac{\zeta}{\bar{\lambda}}\frac{\sinh(x/\bar{\lambda})}{\cosh(x/\bar{\lambda})}\right],
\end{equation}
where, as previously defined, $\bar{\lambda}=\sqrt{2\eta/\xi}$.

The linear stability analysis can thus be performed as before. In the present case, instead of \cref{eq polarity-Fourier-instability}, the equations for the Fourier modes of the polarity perturbations read
\begin{subequations} \label{eq polarity-Fourier-instability-polarized}
\begin{align}
\left(\partial_x^2-q^2\right)\delta\tilde{p}_x=0,\\
\left(\partial_x^2-q^2\right)\delta\tilde{p}_y=0.
\end{align}
\end{subequations}
Respectively, the components of the force balance equation, \cref{eq velocity-Fourier-instability}, reduce to
\begin{subequations} \label{eq velocity-Fourier-instability-polarized}
\begin{gather}
\begin{multlined}
\eta\left(2\partial_x^2-q^2-\frac{1}{\lambda^2}\right)\delta\tilde{v}_x+iq\eta\,\partial_x \delta\tilde{v}_y\\
+\left(T_0/h-2\zeta \partial_x\right)\delta\tilde{p}_x-iq\zeta\, \delta\tilde{p}_y=0,
\end{multlined}
\\
\begin{multlined}
iq\eta\,\partial_x\delta\tilde{v}_x+\eta\left(\partial_x^2-2q^2-\frac{1}{\lambda^2}\right)\delta\tilde{v}_y\\
+\left(T_0/h-\zeta\partial_x\right)\delta\tilde{p}_y=0.
\end{multlined}
\end{gather}
\end{subequations}
Finally, the boundary conditions, previously given by \cref{eq boundary-conditions-Fourier-instability,eq boundary-conditions-stress-Fourier-instability}, now become
\begin{subequations} \label{eq boundary-conditions-Fourier-instability-polarized}
\begin{align}
\delta\tilde{p}_x(0) &= 0,\qquad \delta\tilde{p}_x\left(L\right)=0,\\
\delta\tilde{p}_y(0) &= 0,\qquad \delta\tilde{p}_y\left(L\right)=-iq\,\delta\tilde{L},
\end{align}
\end{subequations} \label{eq boundary-conditions-stress-Fourier-instability-polarized}
\begin{subequations}
\begin{align}
\delta\tilde{v}_x(0) &= 0,\qquad \delta\tilde{\sigma}_{xx}\left(L\right)=\mp \partial_x\sigma_{xx}^0\left(L_0\right)\delta\tilde{L},\\
\delta\tilde{v}_y(0) &= 0,\qquad \delta\tilde{\sigma}_{xy}\left(L\right)=0.
\end{align}
\end{subequations}

\begin{figure}[tb]
\begin{center}
\includegraphics[width=\columnwidth]{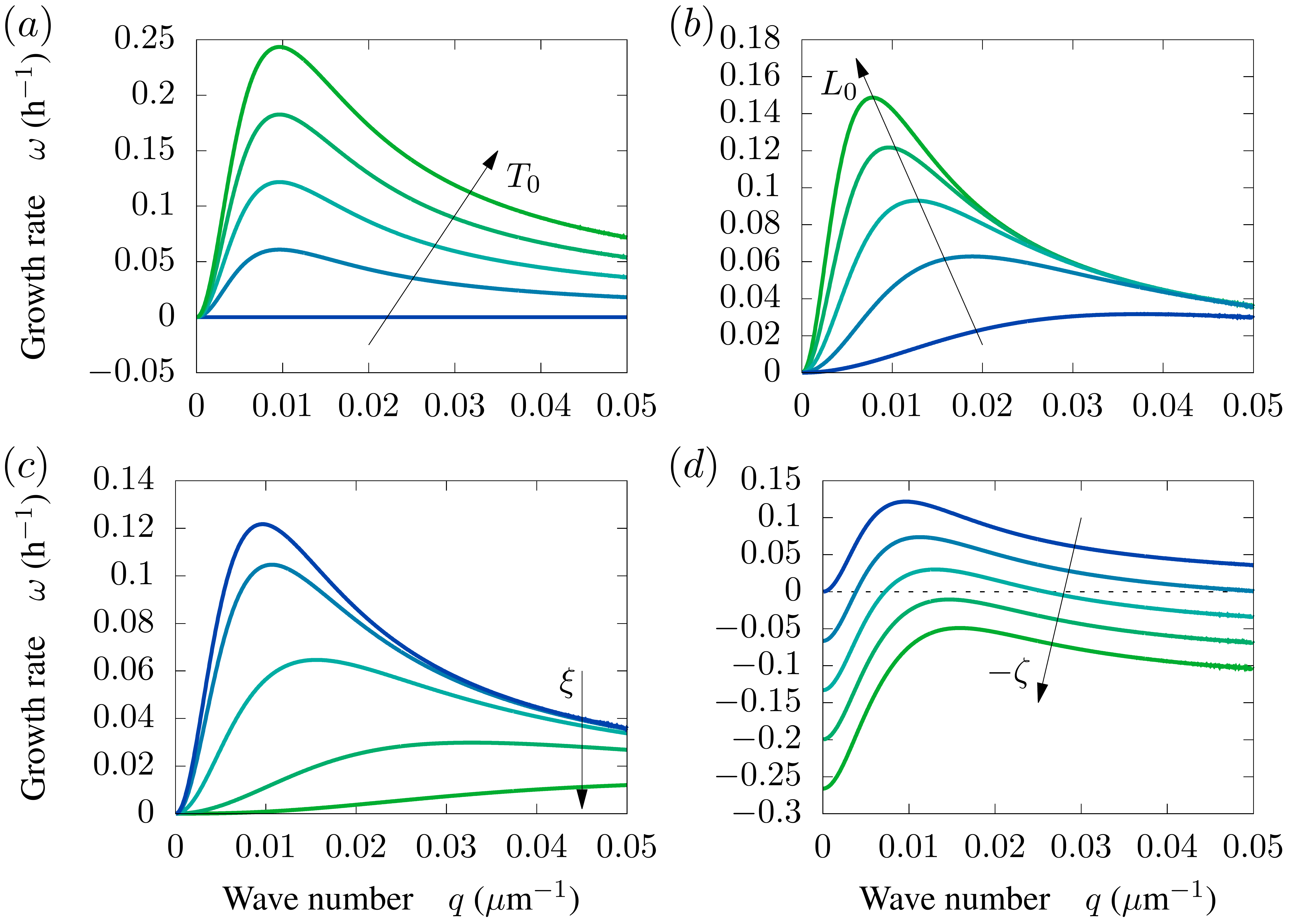}
\caption{Contributions to the instability for a uniformly polarized tissue. Growth rates of front perturbations varying the values of different model parameters. Excluding the varied parameter, other parameter values are in Table I in the Main Text except for $\zeta\rightarrow 0$. (a) Even though the uniform mode ($q=0$) is marginally stable, traction forces destabilize the monolayer front. For this plot, $T_0=0,0.25,0.5,0.75,1$ kPa. (b) Long-range transmission of viscous stresses selects the fastest-growing mode. For this plot $L_0=50,100,150,200,250$ $\mu$m. (c) Cell-substrate friction screens hydrodynamic interactions to limit the wavelength of the fingering pattern. For this plot, $\xi=10^2,5\cdot 10^2,2.5\cdot 10^3,1.25\cdot 10^4,6.25\cdot 10^4$ Pa$\cdot$s/$\mu$m$^2$. (d) Contractility has a stabilizing contribution on the monolayer front. For small contractility, a band of unstable modes at finite wavelength remains. A sufficiently large contractility stabilizes the front. For this plot, $-\zeta=0,1,2,3,4$ kPa.} \label{FigS2}
\end{center}
\end{figure}

Then, \cref{eq polarity-Fourier-instability-polarized,eq velocity-Fourier-instability-polarized} are analytically solved for $\delta\tilde{p}_\alpha(x,q)$ and $\delta\tilde{v}_\alpha(x,q)$. Hence, the growth rate of front perturbations is computed using \cref{eq growth-rate}. The resulting expression is omitted here due to its length. However, the growth rate is plotted in \cref{FigS2}, which shows how it changes under variation of different parameters. From these results, we conclude that, as for tissues with only a polarized boundary layer, uniformly polarized tissues also display an active fingering instability. However, in contrast to the boundary layer case, in the absence of contractility, uniformly polarized tissues spread at a constant velocity $V_0=T_0/(\xi h)$. As a consequence, the $q=0$ mode, is now marginally stable. Nonetheless, all other modes are destabilized by the interplay between active traction forces and viscous stresses. As for the case with an active boundary layer, front perturbations give rise to flow perturbations that destabilize the flat front (\cref{FigS2}a). Hence, the active fingering instability is robust to the presence of bulk polarity in the tissue.

Moreover, the role of hydrodynamic interactions is unchanged with respect to the active boundary layer case. As in that case, the finite range of hydrodynamic interactions, given by the smallest between the monolayer width $L_0$ and the hydrodynamic screening length $\lambda=\sqrt{\eta/\xi}$ determines the wavelength of the most unstable mode. Thus, for polarized tissues, varying the monolayer width $L_0$ and the cell-substrate friction coefficient $\xi$ modifies the growth rate in a similar way as for tissues with only a polarized boundary layer (compare \cref{FigS2}b-c to \cref{fig origins}b-c).

Finally, the effect of the contractility in uniformly polarized tissues is a bit different than in unpolarized tissues. As for tissues with only boundary polarity, contractility has a stabilizing effect on the monolayer front. However, for uniformly polarized tissues, the active contractile stress spans throughout the monolayer. As a consequence, instead of just affecting short-wavelength perturbations, contractility now decreases the growth rate of all perturbation modes. Hence, for small contractility, the longest and shortest-wavelength modes become stable, leaving a band of unstable modes at intermediate wavelengths. Therefore, small contractilities do not abrogate the fingering instability. A sufficiently large contractility, however, is able to stabilize all modes, thus suppressing the fingering instability (\cref{FigS2}d).

\end{document}